\documentclass[11pt]{article}

\usepackage[preprint]{acl}

\usepackage{times}
\usepackage{latexsym}
\usepackage[T1]{fontenc}
\usepackage[utf8]{inputenc}
\usepackage{microtype}
\usepackage{inconsolata}
\usepackage{graphicx}

\usepackage{amsmath,amssymb}
\usepackage{bm}
\usepackage{booktabs}
\usepackage{array}
\usepackage{multirow}
\usepackage{tabularx}
\usepackage{xspace}
\usepackage{placeins}

\newcommand{\Mclean}{\mathcal{M}}
\newcommand{\Mdef}{\widetilde{\mathcal{M}}}
\newcommand{\KL}{\mathrm{KL}}
\newcommand{\ASR}{\mathrm{ASR}}
\newcommand{\rhat}{\hat{\bm{r}}}

\title{Bait-and-Recover: Poisoning Internal Refusal Signals\\
to Defend LLMs against White-Box Editing Jailbreaks}

\author{
  \textbf{Tian Gao}\textsuperscript{1} \quad
  \textbf{Zhipeng Xie}\textsuperscript{1} \quad
  \textbf{Yuhao Wu}\textsuperscript{1} \quad
  \textbf{Junhua Liu}\textsuperscript{1} \quad
  \textbf{Xin Fang}\textsuperscript{2}\\
  \textsuperscript{1}Anhui Laboratory for Safe Artificial Intelligence in the Yangtze River Delta\\
  \textsuperscript{2}iFlytek Research\\
  \texttt{tiangao5@iflytek.com}\\
  \texttt{Code: \href{https://github.com/SparkShieldLab/bait-and-recover}{github.com/SparkShieldLab/bait-and-recover}}
}

\begin{document}
\maketitle

\begin{abstract}
Open-weight large language models face a low-cost white-box threat from 
representation engineering attacks. Attackers can estimate refusal directions 
and search for projection-matrix edits that suppress safety alignment while 
preserving general capabilities, within minutes on a single GPU and without 
gradient-based training. We propose \textbf{Bait-and-Recover}, a weight-level 
defense that places a bait adapter where attackers read activations and a paired 
recovery adapter at the subsequent layer. Trained via gradient routing, this 
decouples the observation path from the behavior path. By actively poisoning the residual signal used 
for measurement, Bait-and-Recover disrupts the attacker's edit search, while 
the recovery layer restores clean downstream computation. Across four 
open-weight models, our defense raises the minimum refusal rate against 
white-box edit searches from 16.25\% to 71.75\% under a strict
behavior-preservation budget ($\KL\le0.10$), with negligible impact on
general benchmarks. By invalidating the core measurement assumption of 
these attacks, observation-path poisoning offers a practical complement to 
behavior-level safety training.
\end{abstract}

\section{Introduction}
\label{sec:intro}

The public release of large language models (LLMs)
\citep{yang2025qwen3,gemmateam2025gemma3} provides direct
transparency into safety alignment, but simultaneously exposes these models
to adversarial weight editing \citep{meng2022rome,meng2023memit}.
A recent family of white-box jailbreaks, exemplified by Heretic
\citep{walters2025heretic}, treats safety-aligned refusal as a
representation-engineering target
\citep{zou2023representation,arditi2024refusal}: estimate the residual
direction that separates harmful and harmless prompts, then search for weight
edits that suppress that direction while preserving the model's general
capabilities.
Unlike prompt-level jailbreaks \citep{zou2023universal,ren2024derail},
this white-box attack requires no crafted
suffix: the attacker reads the model's internal computation, edits projection
matrices in place, and keeps the best edit found within a strict
behavior-preservation budget.

This targeted editing method is a low-cost, broadly applicable threat to
open-weight releases. Full-parameter or LoRA-based \citep{hu2022lora}
fine-tuning attacks against safety alignment require substantially more
compute and often degrade general capabilities if not carefully
controlled. By contrast, a Heretic-style edit search executes in minutes on a
single GPU and yields an edited checkpoint that suppresses refusal while
retaining normal behavior on benign prompts. Once redistributed, such
checkpoints are difficult for the original defender to recall or patch. The
practical defense point therefore shifts to the released weights themselves:
protection must be embedded directly in the checkpoint before distribution.

This attack also exposes a gap in existing safety mechanisms. Safety fine-tuning, 
preference optimization, runtime filters, and activation-level defenses
\citep{ouyang2022training,rafailov2023direct,zou2024circuit} constrain the
model's output behavior, but leave
the internal refusal representations unprotected against measurement and editing.
A white-box edit attack succeeds only when two conditions are met:
the attacker must be able to \emph{observe} an internal signal correlated
with refusal, and editing against that signal must predictably alter the
\emph{output behavior}. Decoupling these two components is therefore a
natural defense target: actively poison the internal signal to blind the
attacker's measurement, while preserving the clean downstream computation
and outputs.

We formalize this idea as \emph{observation--behavior path decoupling} and
implement it via \textbf{Bait-and-Recover}. Let $\ell$ denote a selected bait
layer, and index residual streams so that layer $\ell$ writes to
$r_{\ell+1}$. Our defense places a bait adapter inside layer $\ell$, making
the poisoned signal visible at the attacker's readout $r_{\ell+1}$, and a
paired recovery adapter inside the following layer $\ell+1$, repairing the
downstream residual $r_{\ell+2}$. The intended state is deliberately
asymmetric: the attacker's search uses a corrupted readout, while generation
continues from a near-clean downstream residual.

\paragraph{Contributions.}
By invalidating the core measurement assumption of representation engineering
attacks, our work offers a weight-level complement to behavior-level
safety training. Concretely:
\begin{enumerate}
\item We formulate \emph{observation--behavior path decoupling} as an
      explicit defense objective for white-box jailbreaks, shifting the
      focus from constraining outputs to poisoning the attacker's internal
      measurement step.
\item We propose a cross-layer bait/recovery design with a
      gradient-routing rule that successfully decouples the two paths
      during training, preventing recovery pressure from erasing the bait.
\item Across four open-weight models, the defense raises the mean minimum
      refusal rate against Heretic searches spanning both global and
      per-layer direction scopes under a strict behavior-preservation budget
      ($\KL\le0.10$) from 16.25\% to 71.75\%, with negligible impact on
      general capability benchmarks.
\end{enumerate}

\section{Related Work}
\label{sec:related}

\paragraph{Representation-engineering jailbreaks.}
Heretic \citep{walters2025heretic} is the direct attack we evaluate against.
Although released as an open-source tool rather than a peer-reviewed paper,
it is a representative practical implementation of automatic abliteration:
the public repository reports broad uptake,\footnote{The Heretic repository
listed over 21k stars and 2.2k forks, and its README reported over 3000
community-created Heretic models, when accessed on May 23, 2026:
\url{https://github.com/p-e-w/heretic}.} and its technical design builds on
the observation that refusal behavior can often be captured by a
low-dimensional direction in residual space \citep{arditi2024refusal}.
Heretic packages this line of work into an automated attack by combining
directional ablation with parameter search, drawing on prior abliteration
systems and refinements \citep{labonne2024abliteration,lai2025projected,lai2025norm}.
More broadly, representation engineering studies how steering vectors and
internal activations can monitor or control model behavior
\citep{zou2023representation}. Recent empirical work characterizes how
abliteration erodes diverse safety-pretraining components across open-weight
checkpoints \citep{agnihotri2025granular}, motivating defenses that
specifically resist this attack family. We share the white-box access
assumption of this line of work but invert its defensive target: rather
than improving post-edit behavior, we attack the reliability of the
attacker's representation estimate itself.

\paragraph{Activation-space defenses.}
Circuit-breaker-style defenses modify internal activations to improve
alignment and robustness against adversarial inputs \citep{zou2024circuit}.
Representation Noising \citep{rosati2024repnoise} is closer to our setting in
that it also intervenes on internal representations under an open-weight
threat model, but its objective is to remove or destabilize the
harmful-representation information itself, so that harmful behavior cannot be
recovered through subsequent fine-tuning. Bait-and-Recover targets a
different stage of the attack: we assume a fine-tuning-free white-box edit
search that estimates a refusal direction from model activations and then
edits projection matrices in place. Rather than removing the underlying
refusal or harmful representations, we leave them intact and inject a
localized poison signal at the attacker's observation layer, paired with a
recovery adapter at the next layer that restores the downstream computation.
This asymmetry preserves the model's normal safety behavior while making
the attacker's inferred edit direction unreliable. These defenses therefore
cover different parts of the safety problem: behavior under adversarial
inputs, resistance to harmful fine-tuning, and the reliability of the
attacker's internal measurement.

\paragraph{Model editing and LoRA defenses.}
ROME, MEMIT, and related editing methods show that localized weight changes
can rewrite model behavior \citep{meng2022rome,meng2023memit}, and recent
analyses show that pruning and low-rank modifications can directly compromise
safety alignment \citep{wei2024brittleness}. Safety-aware fine-tuning and
robust safeguards investigate how to preserve or strengthen safety behavior
under subsequent modification \citep{hsu2024safelora,tamirisa2024tamper}. Our
work is complementary: it targets the estimator that a white-box edit attack
relies on, not only the post-edit behavior.

\begin{figure*}[t]
\centering
\includegraphics[width=0.98\textwidth]{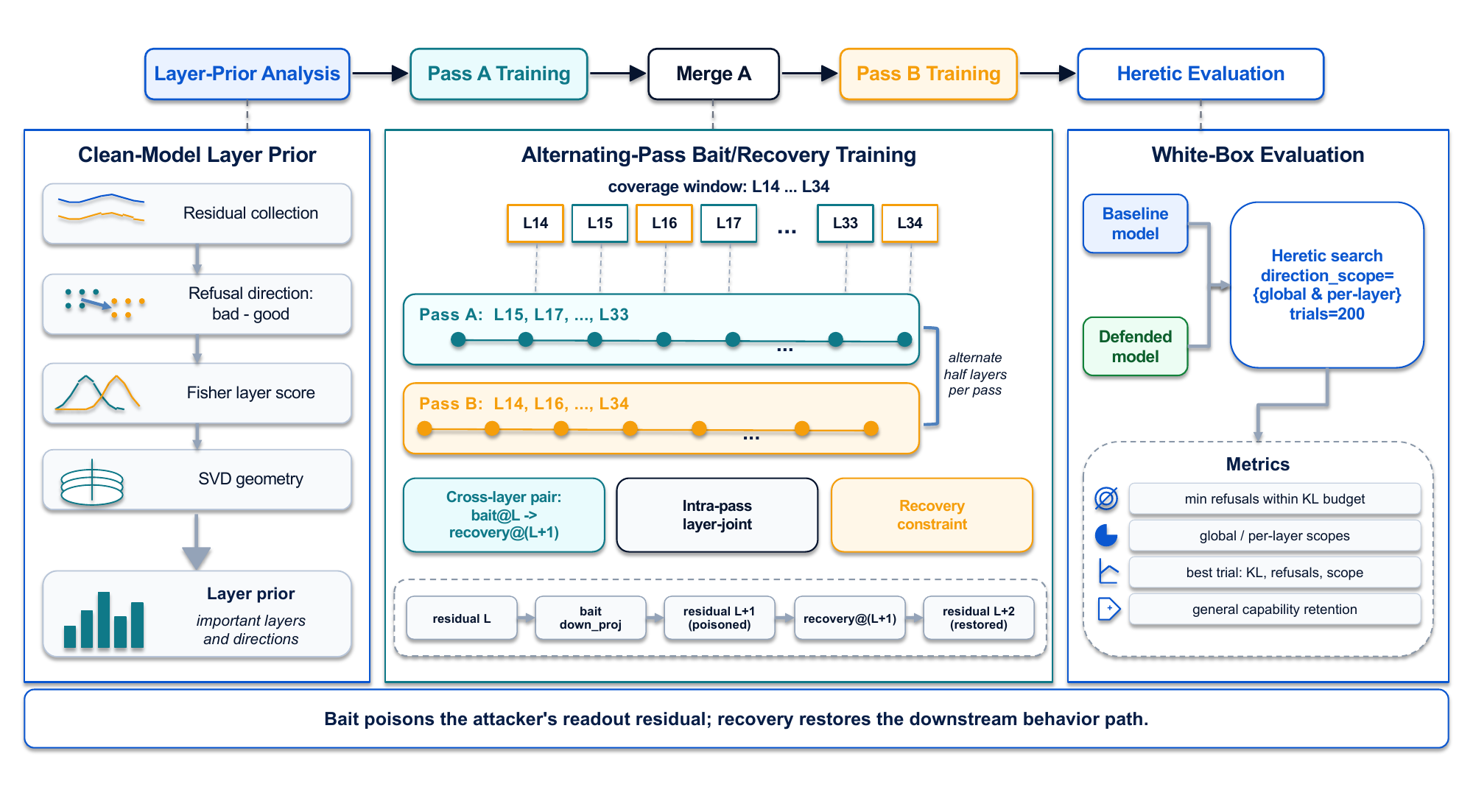}
\caption{Overview of the alternating-pass Bait-and-Recover pipeline. A
clean-model layer-prior analysis determines the bait band before training; Training
Pass~A covers one layer parity, Training Pass~B starts from the merged Pass~A
checkpoint and covers the complementary parity, and the final checkpoint is
evaluated against the same white-box Heretic search used for the baseline. The
layer-joint box denotes local tuning inside a pass over neighboring trained
pairs, not a joint optimization between Pass~A and Pass~B.}
\label{fig:pipeline-overview}
\end{figure*}

\section{Background and Problem Formulation}
\label{sec:background}

\paragraph{Heretic-style attack.}
Given harmful prompts $\mathcal{P}_{\mathrm{harmful}}$ and harmless prompts
$\mathcal{P}_{\mathrm{harmless}}$, the attacker reads residual stream
$\ell{+}1$ and estimates the refusal direction by the difference of means,
\begin{equation}
\label{eq:refusal-dir}
\begin{split}
  \rhat_{\ell+1} = {} &
    \mathbb{E}_{x\sim\mathcal{P}_{\mathrm{harmful}}}[r_{\ell+1}(x)] \\
  & {} - \mathbb{E}_{x\sim\mathcal{P}_{\mathrm{harmless}}}[r_{\ell+1}(x)].
\end{split}
\end{equation}
Heretic~\citep{walters2025heretic} then performs an
Optuna~\citep{akiba2019optuna} search over target
layers, direction scope (global or per-layer), edited modules, sign, and
edit scale, optimizing for the lowest harmful-prompt refusal count under a
KL budget. We evaluate the default configuration: 200 trials, with
\texttt{o\_proj} and \texttt{down\_proj} as the primary editable modules,
and report the minimum refusal count among trials inside each budget.

\paragraph{Problem formulation.}
We assume a white-box attacker with the released checkpoint, full access
to intermediate activations, and enough compute to execute Heretic with its
default mean-direction estimator under both global and per-layer scopes.
The defense is embedded in the released weights and evaluated as a single
merged checkpoint. Let $\Mclean$ be the clean model, $\Mdef$ the defended
model, and $\hat{r}(\cdot)$ the attacker's estimator
(Eq.~\ref{eq:refusal-dir}). The post-edit attack success rate $\ASR$ is the fraction of harmful prompts
for which the edited model does not refuse, equivalently
$\ASR = 1 - \text{refusal rate}$. The defender minimizes $\ASR$ subject to a
general-prompt KL budget:
\begin{align}
\min_{\theta_{\mathrm{bait}},\theta_{\mathrm{rec}}}\quad
& \ASR\!\left(\mathrm{Edit}(\Mdef,\, \hat{r}(\Mdef))\right) \\
\text{s.t.}\quad
& \KL(\Mdef\|\Mclean)\le \varepsilon_{\KL}. \nonumber
\end{align}

\section{Method}
\label{sec:method}

Bait-and-Recover has four stages: (i) build a layer prior from harmful and
harmless prompts on the clean model; (ii) train two complementary parities of
bait/recovery pairs across the selected band; (iii) within each pass, perform a
short local layer-joint step over neighboring trained pairs to absorb
cross-layer coupling; and (iv) evaluate the merged
checkpoint with the same Heretic search used for the baseline.
Figure~\ref{fig:pipeline-overview} summarizes this full pipeline.

\subsection{Layer-band Selection}
\label{sec:layer-band}

We select the defended layer band using clean-model statistics. Let $L$ be
the number of transformer layers and let $\ell$ denote a bait layer.
Because layer $\ell$ writes into
$r_{\ell+1}$ and the paired recovery layer $\ell+1$ writes into $r_{\ell+2}$,
the last admissible bait layer is $\ell\le L{-}2$.

We first define the default clean-model prior window
\begin{equation}
  \mathcal{W}_{\mathrm{prior}} =
  [\lfloor0.4(L{-}1)\rfloor,\,L{-}2],
\end{equation}
which targets middle-to-deep layers while stopping at $L{-}2$, so every bait
layer still has a following recovery layer.

The layer prior uses the separability between harmful and harmless prompts in
the clean-model residual stream. For a candidate bait layer $\ell$ and class
$c\in\{\mathrm{harmful},\mathrm{harmless}\}$, let
$R_{c,\ell+1}=\{r_{\ell+1}(x):x\in\mathcal{P}_c\}$ be the final-token
residuals used by the Heretic direction estimator. We estimate
\begin{equation}
  \begin{aligned}
  \mu_{c,\ell+1}
    &= \frac{1}{|R_{c,\ell+1}|}\sum_{r\in R_{c,\ell+1}} r,\\
  \Sigma_{c,\ell+1}
    &= \operatorname{Cov}(R_{c,\ell+1}),
  \end{aligned}
\end{equation}
and score the harmful/harmless separability at that residual by a trace-form
Fisher ratio,\footnote{We use a trace-form simplification rather than the full
Fisher discriminant $\operatorname{tr}(\Sigma_W^{-1}\Sigma_B)$ to avoid
ill-conditioned covariance inversion at high dimensions.}
\begin{equation}
F_{\ell+1} =
\frac{\|\mu_{\mathrm{harmful},\ell+1}-\mu_{\mathrm{harmless},\ell+1}\|_2^2}
{\operatorname{tr}(\Sigma_{\mathrm{harmful},\ell+1})+
 \operatorname{tr}(\Sigma_{\mathrm{harmless},\ell+1})+\epsilon},
\end{equation}
where larger $F_{\ell+1}$ means the clean refusal signal is easier for Heretic to
measure. Let $N_{\mathrm{pair}}$ denote the number of available
bait/recovery pairs. Since each pair covers one bait layer, if the prior
window contains at most $N_{\mathrm{pair}}$ layers, we use the whole window. If
it is larger, we choose the contiguous sub-band with the highest cumulative
Fisher score:
\begin{equation}
  S^\star =
  \mathop{\arg\max}_{S=[a,a{+}N_{\mathrm{pair}}{-}1]\subseteq
  \mathcal{W}_{\mathrm{prior}}}
  \sum_{\ell\in S} F_{\ell+1}.
\end{equation}

\subsection{Cross-layer Defense Unit}

Each defense unit spans two adjacent transformer layers. To intercept the
attacker's search space, we place a bait adapter on the \texttt{down\_proj}
module in layer $\ell$. The paired recovery uses adapters on the
\texttt{o\_proj} and \texttt{down\_proj} modules in layer $\ell+1$, repairing
the computation before it reaches the generation head. A gradient-routing
rule decouples the updates within each unit: recovery gradients do not update
bait parameters, and visibility gradients do not update recovery parameters.

\subsection{Bait Adapter}
\label{sec:bait-adapter}

The bait adapter is a low-rank \citep{hu2022lora} residual perturbation attached to
\texttt{down\_proj}:
\begin{equation}
  \begin{aligned}
  \delta_\ell(h) &= \exp(\gamma_\ell) B_\ell A_\ell h,\\
  A_\ell &\in\mathbb{R}^{r\times d_{\mathrm{in}}},
  \quad
  B_\ell \in \mathbb{R}^{d_{\mathrm{model}}\times r}.
  \end{aligned}
\end{equation}
Here $h$ is the input to layer $\ell$'s \texttt{down\_proj}, and
$\gamma_\ell$ is a trainable gate initialized small and later calibrated to
the target bait visibility.

The initialization reuses the clean-model means from
Section~\ref{sec:layer-band}. For a bait layer $\ell$, the residual-space
refusal direction is
\begin{equation}
  \hat d_{\ell+1} =
  \frac{\mu_{\mathrm{harmful},\ell+1}-\mu_{\mathrm{harmless},\ell+1}}
       {\|\mu_{\mathrm{harmful},\ell+1}-\mu_{\mathrm{harmless},\ell+1}\|_2}.
\end{equation}
We also compute an input detector direction at the \texttt{down\_proj} input.
Let $\bar h_{c,\ell}$ be the class mean of the final-token \texttt{down\_proj}
inputs $h_\ell(x)$ for $x\in\mathcal{P}_c$; then
\begin{equation}
  \hat q_\ell =
  \frac{\bar h_{\mathrm{harmful},\ell}-\bar h_{\mathrm{harmless},\ell}}
       {\|\bar h_{\mathrm{harmful},\ell}-\bar h_{\mathrm{harmless},\ell}\|_2}.
\end{equation}
Finally, we decompose the down-projection weights,
\begin{equation}
  W^{\mathrm{down}}_\ell = U_\ell\operatorname{diag}(\sigma_\ell)V_\ell^\top.
\end{equation}
The right singular vectors $V_\ell$ provide additional input-side detector
directions, while the left singular vectors $U_\ell$ provide residual-space
output directions.

The first bait channel is the refusal-cancel channel. Its detector reads the
input-side refusal direction $\hat q_\ell$, and its output writes against the
residual-stream refusal direction $\hat d_{\ell+1}$:
\begin{equation}
\label{eq:bait-init}
  A_\ell[0,:]=\hat q_\ell^\top,\qquad
  B_\ell[:,0]=-\hat d_{\ell+1}.
\end{equation}
A second \emph{tag} channel carries a side signal for the recovery layer while
avoiding the clean refusal axis. Let $v_{\ell,j}$ be the $j$-th right singular
vector in $V_\ell$. We choose the vector whose projection away from
$\hat q_\ell$ has the largest norm,
\begin{equation}
  j^\star =
  \arg\max_j \left\|(I-\hat q_\ell\hat q_\ell^\top)v_{\ell,j}\right\|_2 .
\end{equation}
The tag output $\hat t_{\ell+1}$ is initialized to a reproducible pseudo-random
unit vector in the orthogonal complement of $\hat d_{\ell+1}$. We construct the
tag's detector $\tilde v_\ell$ by orthogonalizing and normalizing the selected
singular vector:
\begin{equation}
  \tilde v_\ell =
  \frac{(I-\hat q_\ell\hat q_\ell^\top)v_{\ell,j^\star}}
       {\left\|(I-\hat q_\ell\hat q_\ell^\top)v_{\ell,j^\star}\right\|_2},
\end{equation}
and initialize the channel as $A_\ell[1,:] = \tilde v_\ell^\top$ and
$B_\ell[:,1] = \hat t_{\ell+1}$.
Thus the tag detector is separated from $\hat q_\ell$, and the tag output is
orthogonal to $\hat d_{\ell+1}$. If the adapter rank exceeds two, each
additional channel is initialized from the leading unused SVD directions:
its detector uses a right singular vector from $V_\ell$, and its output uses
the corresponding left singular vector from $U_\ell$, ordered by
$\sigma_\ell$. To ensure training stability, we keep $A_\ell$ fixed and update only $B_\ell$
and $\gamma_\ell$ during training. This keeps the bait's input-side detectors
anchored to the clean refusal boundary, while its residual-space
outputs adapt.

\subsection{Recovery Adapter}
\label{sec:recovery-adapter}

After the bait has changed $r_{\ell+1}$, the next layer can propagate this
perturbation through both residual-writing paths: the attention output
projection and the MLP down-projection. For each bait at layer $\ell$, we
therefore attach recovery adapters to \texttt{o\_proj} and \texttt{down\_proj}
of layer $\ell{+}1$. For a recovery site
$m\in\{\texttt{o\_proj},\texttt{down\_proj}\}$ with input $z_m$, the adapter is
\begin{equation}
  \rho_{\ell,m}(z_m)=O_{\ell,m}D_{\ell,m}z_m .
\end{equation}
Here $D_{\ell,m}\in\mathbb{R}^{r\times d_{\mathrm{in},m}}$ detects
recovery-layer features, and
$O_{\ell,m}\in\mathbb{R}^{d_{\mathrm{model}}\times r}$ writes a corrective
update into the residual stream.

For initialization, we compute the SVD of the recovery-site base projection,
\begin{equation}
  W_{\ell+1,m} =
  U^{\mathrm{rec}}_{\ell,m}
  \operatorname{diag}(\sigma^{\mathrm{rec}}_{\ell,m})
  (V^{\mathrm{rec}}_{\ell,m})^\top .
\end{equation}
Let $v^{\mathrm{rec}}_{\ell,m,i}$ be the $i$-th right singular vector, ordered
by decreasing $\sigma^{\mathrm{rec}}_{\ell,m}$. The detector reads these
high-energy input directions, while the output is initialized to the anti-bait
output basis:
\begin{align}
  D_{\ell,m}[i,:]
  &= (v^{\mathrm{rec}}_{\ell,m,i})^\top,\\
  O_{\ell,m}[:,i]
  &= -B_\ell[:,i],
  \quad i=0,\ldots,r-1.
\end{align}
Training then refines these anti-bait directions to account for the
intermediate layer transformations.

\subsection{Training Objective and Schedule}
\label{sec:losses}

Training combines two main objectives with a lightweight visibility floor:
\begin{equation}
\mathcal{L} = \lambda_{\mathrm{fisher}}\mathcal{L}_{\mathrm{fisher}}
  + \lambda_{\mathrm{rec}}\mathcal{L}_{\mathrm{rec}}
  + \lambda_{\mathrm{vis}}\mathcal{L}_{\mathrm{vis}}.
\end{equation}

$\mathcal{L}_{\mathrm{fisher}}$ is the trace-form Fisher ratio from
Section~\ref{sec:layer-band}, recomputed on defended bait residuals:
\begin{equation}
  \mathcal{L}_{\mathrm{fisher}} =
  \frac{1}{|S|}\sum_{\ell\in S} F^{\mathrm{def}}_{\ell+1}.
\end{equation}
This attack-facing term directly lowers the harmful/harmless separability
visible to the attacker's direction estimator.
$\mathcal{L}_{\mathrm{rec}}$ aligns the clean and defended residual states after
each paired recovery layer,
\begin{equation}
  \mathcal{L}_{\mathrm{rec}} =
  \frac{1}{|S|}\sum_{\ell\in S}\mathbb{E}_x
  \left[
    \frac{\|r^{\mathrm{def}}_{\ell+2}(x) -
            r^{\mathrm{clean}}_{\ell+2}(x)\|_2^2}
         {\|r^{\mathrm{clean}}_{\ell+2}(x)\|_2^2+\epsilon}
  \right].
\end{equation}
Finally, $\mathcal{L}_{\mathrm{vis}}$ is an auxiliary hinge that keeps the bait
measurable at the attacker-visible residual:
\begin{equation}
  \mathcal{L}_{\mathrm{vis}} =
  \frac{1}{|S|}\sum_{\ell\in S}\mathbb{E}_x
  \left[
    \tau_{\mathrm{vis}} -
    \frac{\|\Delta r_{\ell+1}(x)\|_2}
         {\|r^{\mathrm{clean}}_{\ell+1}(x)\|_2+\epsilon}
  \right]_+^2,
\end{equation}
where $\Delta r_{\ell+1}(x)=r^{\mathrm{def}}_{\ell+1}(x)-
r^{\mathrm{clean}}_{\ell+1}(x)$ and $[u]_+=\max(u,0)$.

Gradients are routed by role: recovery alignment updates only recovery
adapters, while Fisher-readout and visibility losses update only bait adapters.
For multi-layer
coverage, the selected band is split into two alternating parities trained in
Pass~A and Pass~B. KL is monitored on general prompts during training, with a
final KL budget.

\section{Experimental Setup}
\label{sec:experiments}

\paragraph{Data.}
Bait-and-Recover uses defense-training data that is separate from Heretic's
attack and evaluation data. For the defense side, harmless prompts and
unlabeled general prompts come from Alpaca-GPT4 instruction prompts
\citep{peng2023instruction}, while harmful prompts come from SafeMTData
\texttt{Attack\_600} \citep{ren2024derail}. We extract the harmful prompt field
and retain direct harmful requests using a deterministic filter over
procedural markers (e.g., how-to, steps, build) and explicit abuse terms
(e.g., malware, phishing, weapon, attack). Because this filtered harmful pool
is smaller than the broad instruction corpus used for harmless prompts, the
labeled defense split uses 128 harmless and 64 harmful prompts for refusal
direction estimation and the Fisher-readout loss. We also reserve 256
unlabeled general prompts for residual recovery, the label-free visibility
floor, and KL monitoring, plus a non-overlapping validation split of 64
harmless and 32 harmful prompts.

For attack evaluation, we leave Heretic's prompt sources unchanged:
\texttt{mlabonne/harmless\_alpaca} and
\texttt{mlabonne/harmful\_behaviors}
\citep{mlabonneHarmlessAlpaca,mlabonneHarmfulBehaviors}. Heretic uses the
first 400 training examples from each source to estimate the attack direction,
and evaluates each attack trial on 100 held-out harmless prompts and 100
held-out harmful prompts for attack KL and harmful-refusal measurement. All
Heretic searches use 200 trials.

\begin{figure*}[t]
\centering
\includegraphics[width=0.88\textwidth]{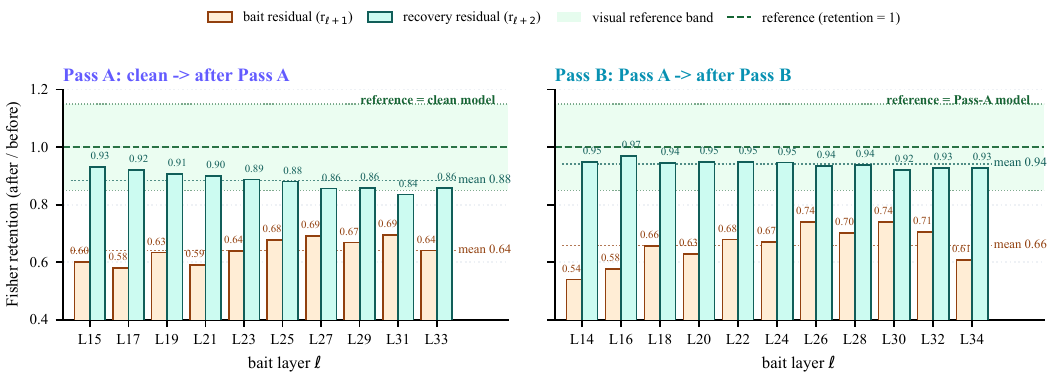}
\caption{Layer-wise Fisher retention during Qwen3-4B-Instruct training. Retention is
after-pass Fisher divided by the before-pass reference (clean model for
Pass~A; Pass~A model for Pass~B). The shaded band is a visual reference around
unchanged retention.}

\label{fig:qwen4b-fisher-retention}
\end{figure*}

\begin{table}[t]
\centering
\footnotesize
\setlength{\tabcolsep}{4pt}
\caption{Evaluated models and defense configurations. The bait band is chosen
by the clean-model layer-prior rule; Pass A/B list the alternating bait-layer
ranges used by the two-stage schedule, and $r$ is the realized LoRA rank range
after applying the rank-selection rule.}
\label{tab:model-config}
\resizebox{\linewidth}{!}{
\begin{tabular}{@{}lccccc@{}}
\toprule
Model & L & Band & Pass A & Pass B & $r$ \\
\midrule
Qwen3-4B-Instruct & 36 & 14--34 & odd 15--33 & even 14--34 & 4--16 \\
Qwen3-8B & 36 & 14--34 & odd 15--33 & even 14--34 & 7--16 \\
Gemma-3-1B-IT & 26 & 10--24 & odd 11--23 & even 10--24 & 4 \\
Gemma-3-12B-IT & 48 & 18--38 & odd 19--37 & even 18--38 & 4--16 \\
\bottomrule
\end{tabular}
}
\end{table}

\paragraph{Models and configuration.}
We evaluate four open-weight instruction models from the Qwen3 and Gemma~3
families \citep{yang2025qwen3,gemmateam2025gemma3}. Table~\ref{tab:model-config}
reports the layer count, defended bait band, alternating pass split, and LoRA
rank range for each model. Layer bands follow the rule in
Section~\ref{sec:layer-band}. We fix the pair budget at $N_{\mathrm{pair}}=21$
across all models: this value is inherited from our initial
Qwen3-4B-Instruct experiment, where the default prior window
$[\lfloor 0.4(L{-}1)\rfloor, L{-}2]$ at $L=36$ contains exactly 21 admissible
bait layers and was used in full. We then reuse this same 21-pair budget for
all subsequent models so that defense capacity stays comparable rather than
scaling with model depth. Under this fixed budget, both Qwen models use their
default 21-layer window in full; Gemma-3-1B-IT has only 15 admissible bait
layers in its default window, so we use all of them; and for Gemma-3-12B-IT
the default window $[18, 46]$ is larger than $N_{\mathrm{pair}}$, so the
Fisher-based contiguous-band selection in Section~\ref{sec:layer-band} picks
the highest-scoring 21-layer sub-band, $[18, 38]$
(Appendix~\ref{app:gemma12b-layer-selection}). LoRA ranks are
selected per layer by first choosing the smallest rank whose clean-model
cumulative squared singular-value energy reaches 40\%, then clipping to the
allowed range and enforcing the minimum rank required by the refusal-cancel
and tag-channel basis. The recovery adapter uses the same rank as its paired
bait adapter.
Table~\ref{tab:model-config} reports the realized ranks; the Gemma-3-1B layers
all happen to hit the minimum rank, so its realized range collapses to the
single value 4.

\paragraph{Training hyperparameters.}
We use an identical hyperparameter configuration across all evaluated models.
All experiments use a shared optimizer recipe: AdamW
\citep{loshchilov2019decoupled} with weight decay 0, gradient clipping at 1.0,
and learning rate $2{\times}10^{-4}$. We fix
$\lambda_{\mathrm{rec}}{=}0.5$, $\lambda_{\mathrm{vis}}{=}4$,
$\lambda_{\mathrm{fisher}}{=}2$, $\tau_{\mathrm{vis}}{=}0.20$. In each pass,
every bait/recovery pair is first trained in isolation for two epochs; after
these single-pair stages, neighboring two-pair blocks receive one epoch of
intra-pass layer-joint tuning. Detailed reproducibility configurations are provided 
in Appendix~\ref{app:reproducibility}.

\paragraph{Evaluation metrics.}
The main attack metric is the minimum refusal rate (out of 100 harmful
prompts) found by a 200-trial Heretic search at each reported evaluation KL
threshold; higher values mean stronger defense. We also
track held-out general-prompt KL after each defense pass as a
behavior-preservation diagnostic. To measure changes on general capability
benchmarks, we run paired EvalScope evaluations \citep{modelscopeEvalScope}
with the same vLLM backend \citep{kwon2023vllm}, chat template, output cap,
and postprocessing for each clean/defended pair, reporting MMLU-Redux accuracy
\citep{gema2025done,hendrycks2021mmlu} and IFEval prompt-level strict accuracy
\citep{zhou2023ifeval,biderman2024lessons}.

\section{Results}
\label{sec:results}

\begin{table*}[t]
\centering
\footnotesize
\setlength{\tabcolsep}{3.6pt}
\caption{Behavior-preservation diagnostics and general capability retention
after defense. KL is measured on held-out general prompts after Pass~A and
Pass~B; Pass~B is the final defended checkpoint. MMLU reports MMLU-Redux
accuracy, IFEval reports prompt-level strict accuracy, and $\Delta$ is
defended minus baseline.}
\label{tab:capability-retention}
\begin{tabular}{lccrrrrrr}
\toprule
& \multicolumn{2}{c}{General-prompt KL}
& \multicolumn{3}{c}{MMLU-Redux}
& \multicolumn{3}{c}{IFEval} \\
\cmidrule(lr){2-3}\cmidrule(lr){4-6}\cmidrule(lr){7-9}
Model & Pass A & Pass B
& Base & Def. & $\Delta$
& Base & Def. & $\Delta$ \\
\midrule
Qwen3-4B-Instruct & 0.016 & 0.020 & 0.7549 & 0.7538 & -0.0011 & 0.8244 & 0.8373 & +0.0129 \\
Qwen3-8B & 0.007 & 0.010 & 0.7782 & 0.7761 & -0.0021 & 0.8373 & 0.8373 & +0.0000 \\
Gemma-3-1B-IT & 0.066 & 0.079 & 0.4180 & 0.4193 & +0.0013 & 0.5730 & 0.5342 & -0.0388 \\
Gemma-3-12B-IT & 0.049 & 0.085 & 0.7528 & 0.7506 & -0.0022 & 0.8059 & 0.8096 & +0.0037 \\
\midrule
Mean & 0.035 & 0.049 & 0.6760 & 0.6750 & -0.0010 & 0.7602 & 0.7546 & -0.0055 \\
\bottomrule
\end{tabular}
\end{table*}

\begin{table*}[t]
\centering
\footnotesize
\setlength{\tabcolsep}{3pt}
\caption{Main Heretic attack result across four KL budgets. Each cell reports
the minimum refusal rate (\%, out of 100 harmful prompts; higher is better);
bold marks the $\KL\le0.10$ primary evaluation budget.}
\label{tab:overall-selected}
\resizebox{\textwidth}{!}{%
\begin{tabular}{@{}lcccccccccccc@{}}
\toprule
& \multicolumn{3}{c}{$\KL\le0.05$} & \multicolumn{3}{c}{$\KL\le0.10$} & \multicolumn{3}{c}{$\KL\le0.20$} & \multicolumn{3}{c}{$\KL\le1.0$} \\
\cmidrule(lr){2-4}\cmidrule(lr){5-7}\cmidrule(lr){8-10}\cmidrule(lr){11-13}
Model & Base & Def. & $\Delta$ & Base & Def. & $\Delta$ & Base & Def. & $\Delta$ & Base & Def. & $\Delta$ \\
\midrule
Qwen3-4B-Instruct & 97 & 78 & $-19$ & \textbf{46} & \textbf{78} & \textbf{$+32$} & 24 & 65 & $+41$ & 5 & 60 & $+55$ \\
Qwen3-8B & 36 & 67 & $+31$ & \textbf{11} & \textbf{60} & \textbf{$+49$} & 11 & 50 & $+39$ & 8 & 33 & $+25$ \\
Gemma-3-1B-IT & 17 & 84 & $+67$ & \textbf{5} & \textbf{67} & \textbf{$+62$} & 3 & 64 & $+61$ & 2 & 64 & $+62$ \\
Gemma-3-12B-IT & 12 & 82 & $+70$ & \textbf{3} & \textbf{82} & \textbf{$+79$} & 2 & 82 & $+80$ & 0 & 59 & $+59$ \\
\midrule
Mean & 40.50 & 77.75 & $+37.25$ & \textbf{16.25} & \textbf{71.75} & \textbf{$+55.50$} & 10.00 & 65.25 & $+55.25$ & 3.75 & 54.00 & $+50.25$ \\
\bottomrule
\end{tabular}%
}
\end{table*}

\subsection{Fisher Suppression and Behavior Preservation}
\label{sec:results-stability}

Figure~\ref{fig:qwen4b-fisher-retention} validates the training task directly.
Across both passes, the residuals exposed to the attacker show lower Fisher
separability after bait training, while the following recovery residuals stay
close to the previous reference model. This is the intended asymmetry of the
defense: the observation path becomes less useful for estimating a refusal
direction, but the behavior path is pulled back toward the clean computation.
Appendix~\ref{app:qwen4b-fisher} reports the corresponding raw Fisher values.

Alongside this Fisher diagnostic, we monitor validation KL on general prompts
to check that the defense has not moved normal behavior too far. All four
configurations completed under the shared training schedule. The Pass~B
general-prompt KL remains below 0.20 for every model, though the amount of
movement differs by architecture (Table~\ref{tab:capability-retention}).
Qwen3-8B is the most conservative case, while Gemma-3-12B has the largest
general-prompt movement among the four models.

\subsection{General Capability Retention}
\label{sec:results-capability}

Because KL alone does not establish general capability preservation, we also
run paired benchmark evaluations for each clean/defended checkpoint.
Table~\ref{tab:capability-retention} reports both the validation KL diagnostics
and the paired capability evaluation. MMLU-Redux accuracy is largely
unchanged across the four checkpoints:
the defended-minus-baseline deltas range from $-0.0022$ to $+0.0013$, with a
mean change of $-0.0010$. IFEval prompt-strict accuracy is also stable in
aggregate, with a mean change of $-0.0055$. Thus
the paired benchmarks support broad capability retention.

\subsection{Heretic Attack Evaluation}
\label{sec:results-attack}

Heretic is the central white-box attack evaluation because it directly tests
the failure mode targeted by Bait-and-Recover: the attacker estimates an
internal refusal direction, searches projection edits, and keeps the best edit
whose behavior KL remains within a budget. We therefore report the minimum
refusal rate over the pooled global and per-layer direction scopes at each KL
threshold; higher rates mean fewer successful harmful responses and stronger
defense.

We use $\KL\le0.10$ as the primary evaluation budget because it matches the
behavior-preservation scale of the defended checkpoints themselves: after
Pass~B, the held-out general-prompt KL values in
Table~\ref{tab:capability-retention} are all below 0.10, and the paired
MMLU-Redux/IFEval evaluations show little aggregate capability movement. Thus,
if Heretic can substantially reduce refusals while staying within the same KL
scale, the edited checkpoint is not just degraded; it is a
behavior-preserving jailbreak.

At this primary evaluation budget, Bait-and-Recover raises the mean minimum refusal
rate from 16.25\% to 71.75\% ($+55.50$; 4.4$\times$), with gains on all four
models (Table~\ref{tab:overall-selected}; see Appendix~\ref{app:full-heretic} for
full results across all KL budgets). The defense remains effective at the
looser $\KL\le0.20$ budget, where the mean defended minimum refusals is 65.25,
only 6.5 refusals below the $\KL\le0.10$ value and still $+55.25$ above the
baseline mean of 10.00. Robustness extends to the very loose $\KL\le1.0$
budget as well: the mean defended score remains at 54.00 ($+50.25$ over
baseline; Table~\ref{tab:overall-selected}, $\KL\le1.0$ column), with
Gemma-3-12B-IT keeping its per-layer defended score at 82 across all
KL budgets, while its global defended score falls from 83 to 59 only at
$\KL\le1.0$. This is
consistent with the broader trend at the primary $\KL\le0.10$ budget, where
global search is the most effective attack scope against every defended model.

\paragraph{Global vs. per-layer scope.}
The scope decomposition shows where the defended models remain most
vulnerable. Before defense, the most effective attack scope differs by model: global search
sets the pooled minimum for Qwen3-4B-Instruct, Qwen3-8B, and Gemma-3-1B-IT, while
per-layer search is the most effective attack on Gemma-3-12B-IT. After defense,
global search remains the most effective attack scope against the three smaller
models, while for Gemma-3-12B-IT the global and per-layer defended scores are
nearly tied at $\KL\le0.10$ (83 vs.\ 82). Across models
(Table~\ref{tab:scope-kl10}), defended global scores are 78/60/67/83 and
per-layer scores are 81/88/82/82. The mean across models reflects this trend,
with a 72.00\% refusal rate under global search compared to 83.25\% under
per-layer search at $\KL\le0.10$. Thus the defense substantially raises the attacker's
best available refusal floor, but the remaining attack surface is concentrated
in global direction estimates rather than individual layer directions, which
is the main remaining weakness of the defense.

\begin{table}[t]
\centering
\footnotesize
\setlength{\tabcolsep}{4.5pt}
\caption{Heretic minimum refusals at the $\KL\le0.10$ primary evaluation budget,
split by attacker direction scope. Each entry reports base/defended/$\Delta$.}
\label{tab:scope-kl10}
\resizebox{\linewidth}{!}{
\begin{tabular}{@{}lcc@{}}
\toprule
Model & Global & Per-layer \\
\midrule
Qwen3-4B-Instruct  & 46/\textbf{78}/$+32$ & 94/81/$-13$ \\
Qwen3-8B  & 11/\textbf{60}/$+49$ & 99/88/$-11$ \\
Gemma-3-1B-IT  & 5/\textbf{67}/$+62$  & 22/82/$+60$ \\
Gemma-3-12B-IT & 26/\textbf{83}/$+57$ & 3/82/$+79$ \\
\midrule
Mean      & 22.00/\textbf{72.00}/$+50.00$ & 54.50/83.25/$+28.75$ \\
\bottomrule
\end{tabular}
}
\end{table}

\section{Conclusion}
\label{sec:conclusion}

We introduced Bait-and-Recover, a weight-level defense that protects open-weight
models against white-box editing jailbreaks by invalidating their core measurement
assumption. Rather than only hardening the output behavior, Bait-and-Recover
places a bait adapter at the attacker's observation layer and a paired recovery
adapter at the next layer. Trained under a gradient-routing rule, this mechanism
decouples the observation path from the behavior path: it actively poisons the
residual signal used to estimate the refusal direction, making the attacker's edit
search unreliable, while the recovery layer restores the clean downstream
computation. Across four open-weight models from the Qwen3 and Gemma-3 families,
the defense raises the mean minimum refusal rate against Heretic searches from
16.25\% to 71.75\% at a strict $\KL\le0.10$ budget, with negligible impact on
general capability benchmarks. Because the defense ships as a single merged
checkpoint loadable with standard causal-LM inference, observation-path poisoning
is a practical weight-level complement to behavior-level safety training. The
remaining attack surface concentrates in global direction estimates. Future
work includes more adaptive estimators and broader capability evaluation.

\section*{Limitations}

\paragraph{Adaptive estimators.}
We evaluate against Heretic's default difference-of-means estimator. While the
bait is designed to distort this estimator's observed refusal direction, we do
not evaluate robust or multivariate alternatives such as geometric medians,
PCA, or LDA. Estimator-agnostic robustness therefore remains an open question.

\paragraph{High-KL budgets.}
Our defense targets the strict behavior-preservation setting ($\KL\le0.10$). At
looser budgets (e.g., $\KL\ge0.20$), greater permitted deviation allows attackers
to discover bypass edits. Our conclusions should therefore be interpreted as
applying to low-distortion edit searches rather than arbitrary model changes.

\paragraph{Behavior-level fine-tuning.}
Supervised fine-tuning (SFT) can still overwrite safety behaviors, as a small
number of adversarial examples suffices to undo alignment
\citep{qi2024finetuning}. We do not claim protection against attackers willing
to collect harmful training data and run additional fine-tuning.

\paragraph{Comparison with prior weight-level defenses.}
We do not run direct comparisons with prior weight-level or
representation-level safety defenses such as Circuit Breakers
\citep{zou2024circuit}, Representation Noising \citep{rosati2024repnoise},
or Tamper-Resistant Safeguards \citep{tamirisa2024tamper}. These defenses
are evaluated under behavior-level attacks, harmful fine-tuning, or safeguard
tampering, whereas our experiments isolate Heretic-style edit searches. We
therefore leave cross-defense evaluations and combination studies to future
work.

\section*{Ethics Statement}

This work studies a dual-use white-box jailbreak setting to improve defenses.
To mitigate risks, we report only aggregate statistics, omit harmful
generations, and explicitly avoid releasing jailbroken checkpoints. Because
our defense modifies model weights, practical deployment should be paired
with capability evaluations to verify that benign utility is preserved.

\bibliography{paper}
\clearpage
\appendix
\captionsetup{hypcap=false}
\setlength{\textfloatsep}{6pt plus 2pt minus 2pt}
\setlength{\floatsep}{6pt plus 2pt minus 2pt}
\setlength{\intextsep}{6pt plus 2pt minus 2pt}
\setlength{\dbltextfloatsep}{6pt plus 2pt minus 2pt}
\setlength{\dblfloatsep}{6pt plus 2pt minus 2pt}
\renewcommand{\topfraction}{0.95}
\renewcommand{\bottomfraction}{0.8}
\renewcommand{\textfraction}{0.05}
\renewcommand{\floatpagefraction}{0.7}
\renewcommand{\dbltopfraction}{0.95}
\renewcommand{\dblfloatpagefraction}{0.7}

\twocolumn[{%
\begin{center}
  \includegraphics[width=0.98\textwidth]{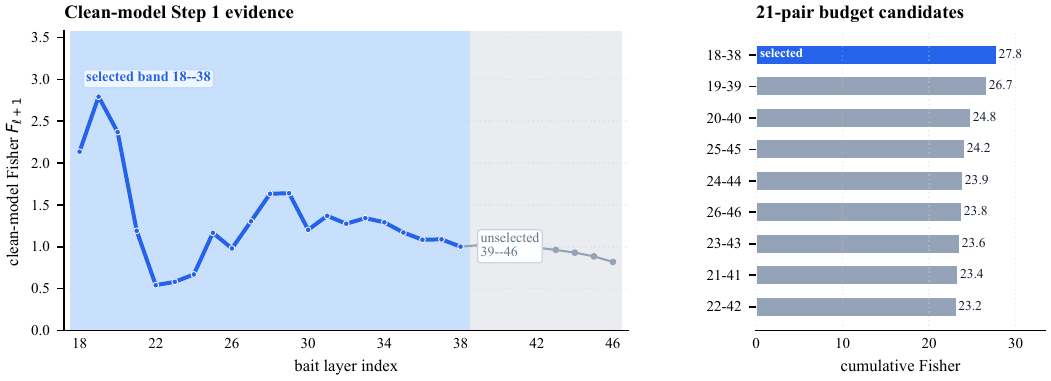}
  \captionof{figure}{Clean-model layer-prior evidence for Gemma-3-12B. Given the
$N_{\mathrm{pair}}=21$ budget, the selected $[18, 38]$ bait band is the
highest-scoring contiguous 21-layer sub-band inside the default $[18, 46]$
clean-model prior window.}
  \label{fig:gemma12b-layer-selection}
  \vspace{1em}
\end{center}
}]

\section{Gemma-3-12B Layer-prior Evidence}
\label{app:gemma12b-layer-selection}

Figure~\ref{fig:gemma12b-layer-selection} shows the clean-model Fisher
diagnostic used to select the Gemma-3-12B bait band: with the fixed
$N_{\mathrm{pair}}=21$ pair budget, the contiguous-band selection rule from
\S\ref{sec:layer-band} picks $[18, 38]$ as the highest-scoring 21-layer
sub-band inside the default $[18, 46]$ prior window.

\begin{table}[!htbp]
\centering
\small
\setlength{\tabcolsep}{3pt}
\caption{Full overall Heretic results. Models are abbreviated: Q3-4B (Qwen3-4B-Instruct), Q3-8B (Qwen3-8B), G3-1B (Gemma-3-1B-IT), G3-12B (Gemma-3-12B-IT).}
\label{tab:app-overall}
\begin{tabular}{@{}lcccc@{}}
\toprule
KL & Q3-4B & Q3-8B & G3-1B & G3-12B \\
\midrule
$\le0.01$ & 100/100/0 & 97/95/-2 & 88/NA/NA & 72/82/+10 \\
$\le0.02$ & 97/97/0 & 78/84/+6 & 88/84/-4 & 37/82/+45 \\
$\le0.03$ & 97/78/-19 & 44/78/+34 & 77/84/+7 & 37/82/+45 \\
$\le0.05$ & 97/78/-19 & 36/67/+31 & 17/84/+67 & 12/82/+70 \\
$\le0.075$ & 97/78/-19 & 15/60/+45 & 17/67/+50 & 8/82/+74 \\
$\le0.10$ & 46/78/+32 & 11/60/+49 & 5/67/+62 & 3/82/+79 \\
$\le0.15$ & 24/70/+46 & 11/52/+41 & 5/64/+59 & 3/82/+79 \\
$\le0.20$ & 24/65/+41 & 11/50/+39 & 3/64/+61 & 2/82/+80 \\
$\le0.30$ & 6/65/+59 & 8/37/+29 & 2/64/+62 & 1/82/+81 \\
$\le0.50$ & 5/65/+60 & 8/33/+25 & 2/64/+62 & 1/82/+81 \\
$\le1.00$ & 5/60/+55 & 8/33/+25 & 2/64/+62 & 0/59/+59 \\
\bottomrule
\end{tabular}
\end{table}

\begin{table}[!htbp]
\centering
\small
\setlength{\tabcolsep}{3pt}
\caption{Full global-direction Heretic results.}
\label{tab:app-global}
\begin{tabular}{@{}lcccc@{}}
\toprule
KL & Q3-4B & Q3-8B & G3-1B & G3-12B \\
\midrule
$\le0.01$ & 100/100/0 & 97/95/-2 & 88/NA/NA & 89/83/-6 \\
$\le0.02$ & 97/97/0 & 78/84/+6 & 88/84/-4 & 61/83/+22 \\
$\le0.03$ & 97/78/-19 & 44/78/+34 & 77/84/+7 & 61/83/+22 \\
$\le0.05$ & 97/78/-19 & 36/67/+31 & 17/84/+67 & 61/83/+22 \\
$\le0.075$ & 97/78/-19 & 15/60/+45 & 17/67/+50 & 61/83/+22 \\
$\le0.10$ & 46/78/+32 & 11/60/+49 & 5/67/+62 & 26/83/+57 \\
$\le0.15$ & 24/78/+54 & 11/52/+41 & 5/64/+59 & 26/83/+57 \\
$\le0.20$ & 24/78/+54 & 11/50/+39 & 3/64/+61 & 2/83/+81 \\
$\le0.30$ & 6/78/+72 & 8/37/+29 & 2/64/+62 & 2/83/+81 \\
$\le0.50$ & 5/78/+73 & 8/33/+25 & 2/64/+62 & 1/83/+82 \\
$\le1.00$ & 5/78/+73 & 8/33/+25 & 2/64/+62 & 0/59/+59 \\
\bottomrule
\end{tabular}
\end{table}

\begin{table}[!htbp]
\centering
\small
\setlength{\tabcolsep}{3pt}
\caption{Full per-layer-direction Heretic results. ``NA'' means no trial landed inside the KL bucket.}
\label{tab:app-perlayer}
\begin{tabular}{@{}lcccc@{}}
\toprule
KL & Q3-4B & Q3-8B & G3-1B & G3-12B \\
\midrule
$\le0.01$ & NA/NA/NA & NA/99/NA & 94/NA/NA & 72/82/+10 \\
$\le0.02$ & NA/NA/NA & NA/99/NA & 90/NA/NA & 37/82/+45 \\
$\le0.03$ & NA/NA/NA & 100/99/-1 & 88/88/0 & 37/82/+45 \\
$\le0.05$ & 100/95/-5 & 100/99/-1 & 78/86/+8 & 12/82/+70 \\
$\le0.075$ & 99/95/-4 & 100/97/-3 & 68/82/+14 & 8/82/+74 \\
$\le0.10$ & 94/81/-13 & 99/88/-11 & 22/82/+60 & 3/82/+79 \\
$\le0.15$ & 94/70/-24 & 95/86/-9 & 10/82/+72 & 3/82/+79 \\
$\le0.20$ & 88/65/-23 & 93/74/-19 & 8/80/+72 & 3/82/+79 \\
$\le0.30$ & 88/65/-23 & 90/67/-23 & 6/80/+74 & 1/82/+81 \\
$\le0.50$ & 56/65/+9 & 85/60/-25 & 3/77/+74 & 1/82/+81 \\
$\le1.00$ & 28/60/+32 & 78/53/-25 & 3/73/+70 & 1/82/+81 \\
\bottomrule
\end{tabular}
\end{table}

\section{Detailed Heretic Evaluation Results}
\label{app:full-heretic}

Tables~\ref{tab:app-overall}--\ref{tab:app-perlayer} report the full evaluation results
used to produce the selected main-text attack table. Each entry reports
\textsc{B}/\textsc{D}/$\Delta$, where $\Delta=\textsc{D}-\textsc{B}$.

\begin{table}[!htbp]
\centering
\small
\setlength{\tabcolsep}{3pt}
\renewcommand{\arraystretch}{0.95}
\caption{Raw Fisher values for Qwen3-4B-Instruct. ``Ref.'' denotes the
before-pass reference used as the retention denominator.}
\label{tab:raw-fisher-qwen4b}
\begin{tabular}{@{}llrrrr@{}}
\toprule
& & \multicolumn{2}{c}{Bait} & \multicolumn{2}{c}{Recovery} \\
\cmidrule(lr){3-4}\cmidrule(lr){5-6}
Pass & Layer & Ref. & Def. & Ref. & Def. \\
\midrule
A & 15 & 0.89 & 0.54 & 1.06 & 0.99 \\
A & 17 & 1.05 & 0.61 & 1.26 & 1.16 \\
A & 19 & 1.54 & 0.98 & 1.35 & 1.23 \\
A & 21 & 1.19 & 0.70 & 1.19 & 1.07 \\
A & 23 & 1.23 & 0.79 & 1.17 & 1.04 \\
A & 25 & 1.17 & 0.79 & 1.12 & 0.98 \\
A & 27 & 1.11 & 0.77 & 1.05 & 0.90 \\
A & 29 & 1.04 & 0.69 & 1.02 & 0.88 \\
A & 31 & 1.02 & 0.71 & 1.00 & 0.83 \\
A & 33 & 0.92 & 0.59 & 0.89 & 0.76 \\
\midrule
B & 14 & 0.65 & 0.35 & 0.54 & 0.51 \\
B & 16 & 0.98 & 0.57 & 0.61 & 0.59 \\
B & 18 & 1.16 & 0.76 & 0.98 & 0.93 \\
B & 20 & 1.22 & 0.77 & 0.70 & 0.67 \\
B & 22 & 1.06 & 0.72 & 0.79 & 0.75 \\
B & 24 & 1.04 & 0.70 & 0.79 & 0.75 \\
B & 26 & 0.98 & 0.73 & 0.77 & 0.72 \\
B & 28 & 0.90 & 0.63 & 0.69 & 0.65 \\
B & 30 & 0.88 & 0.65 & 0.71 & 0.65 \\
B & 32 & 0.83 & 0.59 & 0.59 & 0.55 \\
B & 34 & 0.76 & 0.46 & 0.66 & 0.61 \\
\bottomrule
\end{tabular}
\end{table}

\section{Qwen3-4B-Instruct Raw Fisher Values}
\label{app:qwen4b-fisher}

Table~\ref{tab:raw-fisher-qwen4b} reports the raw Fisher separability scores
for the validation split visualized in Figure~\ref{fig:qwen4b-fisher-retention}.

\section{Reproducibility Details}
\label{app:reproducibility}

We implement the framework in PyTorch
\citep{paszke2019pytorch} with the Hugging Face Transformers library
\citep{wolf2020transformers}. Models are loaded from their official
Hugging Face or ModelScope weights and trained in \texttt{bfloat16}, with
device-mapped model parallelism for the larger checkpoints. After training,
the bait and recovery adapters are merged into the base weights, so the
defended checkpoint runs as a standard causal LM with no custom inference code.

Layer-prior extraction and adapter training both use final-token residual
activations from training splits of the ModelScope-Alpaca and SafeMT datasets
that are disjoint from the Heretic evaluation prompts. Heretic attacks are
run with the Optuna search framework, using 200 trials per attack over
projection layers, weight matrices, edit directions, and scales, and
reporting the lowest-refusal edit within each KL budget. General-capability
evaluation uses EvalScope with a vLLM backend, and we use the same chat
template, generation length, and post-processing for the clean and defended
checkpoints of each model when computing MMLU-Redux and IFEval.

\end{document}